# Wellformedness Properties in Euler Diagrams: An Eye Tracking Study for Visualisation Evaluation

**Mithileysh Sathiyanarayanan**
University of Brighton
United Kingdom
M.Sathiyanarayanan@brighton.ac.uk

**Tobias Mulling**
University of Brighton
United Kingdom
T.Mulling@brighton.ac.uk

**ABSTRACT**
In the field of information visualisation, Euler diagrams are an important tool used in various application areas such as engineering, medicine and social analysis. To effectively use Euler diagrams, some of the wellformedness properties needs to be avoided, as they are considered to reduce user comprehension. From the previous empirical studies, we know some properties are swappable but there is no clear justification which property would be the best to use. In this paper, we considered two main wellformedness properties (duplicated curve labels and disconnected zones) to test which among the two affect user comprehension the most, based on the task performance (accuracy and response time), preference and eye movements of the users. Twelve participants performed three different types of tasks with nine diagrams of each property (so, in total eighteen diagrams) and the results showed that duplicated curve labels property slows down and trigger extra eye movements, causing delays for the tasks. Though there is no significant difference in the accuracy but the insights obtained from the response time, preference and eye movements will be useful for software developers on the optimal way to visualise Euler diagrams in real world application areas.

**ACM Classification Keywords**
H.5.m. Information interfaces and presentation (e.g., HCI): Miscellaneous

## INTRODUCTION
Euler diagrams are also called as set-based diagrams or curve-based diagrams or region-based diagrams which are widely used in many application areas because they represent set intersections, set disjointness and set containments. This allows users to understand which sets the item belong to, and which does not. From the literature, an Euler diagram is defined to be a collection of labeled curves that are closed. A segment connected in the plane that has a curve boundary is called as minimal region and a set of minimal regions is called a zone.

Flower and Howse [5] defined the notion of wellformedness properties that relate to relationships between curves and regions in the diagram. The six properties were defined based on the drawing methods that produced bad layouts. In order to produce aesthetically pleasing diagrams, Rodgers et al. [7] conducted an empirical study considering all the six wellformedness properties and these properties have been proved to affect the understanding of a diagram. The properties are[1]

1. Non-simple curves: a self intersecting curve should be avoided.
2. Duplicated curve labels: curves having same label should be avoided.
3. Concurrency between curves: curves running along same borders should be avoided.
4. N-point between curves: curves crossing at a same point to be avoided.
5. Brushing point between curves: curves touching each other should be avoided.
6. Edges disconnecting zones: each zone should be a minimal region.

After the properties were defined, empirical studies were conducted. We have contradicting results from the empirical studies conducted by Rodgers et al. [7], Fish et al. [4]. Riche and Dwyer [6].

1. An empirical study conducted by Rodgers et al. [7] reveals two phase results. An initial study revealed concurrency and disconnected zones have high significant effect compared to brushing points, n-points, non-simple curves and duplicated curve labels, whereas the second study found duplicated curve labels have significant effect.
2. An empirical study conducted by Fish et al. [4] did not consider the duplicated curve labels property and the study revealed concurrency, disconnected zones and non-simple curves have less significant effect compared to brushing points and n-points.
3. An empirical study conducted by Riche and Dwyer [6] revealed disconnected zones and concurrency are preferred by the users.

---
[1] More detailed explanation with examples are given in [5], [8], [9] and [10].

These contradicting results have not considered participants eye movements. So, our main interest lies in comparing only duplicated curve labels and disconnected zones using eye tracking approach as shown in Figure 1. In the left-hand diagram, curve label C is duplicated but the curves are circular in shape whereas in the right-hand diagram, curve label C is not duplicated but the curve shape is ellipse and it disconnects the region AB. So, we have a strong motivation for discovering which among the two properties affect user understanding the most. The research question this paper attempts to address is ``Which wellformedness properties (disconnected zones and duplicated curve labels) affect user comprehension the most?". The main aim of the study in this paper is to use eye tracking to test the effects of the two wellformedness properties based on eye movements, task performance and preference.

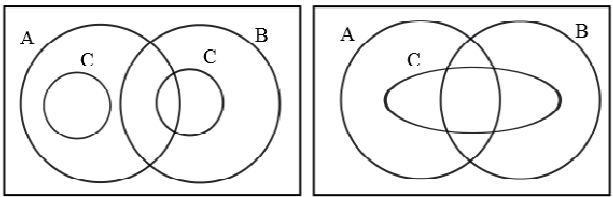

**Figure 1. (a) Curve label C is duplicated but the curves are circular in shape and no disconnection (b) Curve label C is not duplicated but the curve is ellipse in shape and disconnecting.**

Twelve participants answered a series of eighteen questions where each property was asked nine times. Participants' ability was determined by the number of answers being correct and how well they observe and concentrate on the targeted areas in the diagrams to get the right answers. In this study, we considered a within-group approach to compare two wellformedness properties and which among the two affect users' comprehension the most. We employ a non-invasive eye-tracker (Tobii X2-60) to record participants' eye movements during each diagram solving task. The data collected included: participants' fixation time, fixation counts, and scan paths of the critical areas of each diagram.
The results showed that duplicated curve labels property slows down and trigger extra eye movements, causing delays for the tasks. The insights obtained are discussed in order to consider which property can be relaxed while generating the diagrams.

**EXPERIMENTAL DESIGN**
The experiment is a controlled study conducted within a group to test the effects of the two wellformedness properties (disconnected zones and duplicated curve labels) based on eye movements and task performance. Twelve participants from the University of Brighton, UK volunteered to take part in the study who had knowledge about computers and had no vision defects[2]. Few participants were familiar with Euler diagrams and Venn diagrams. An initial training of the diagrams with questions were provided so that the participants get familiar with the diagrams, questions and the complete study process.
**Usability Room**
The usability room had a good ventilation and a dedicated computer with an eye tracking system. Participants were provided with comfortable chairs and were advised not to touch the monitor or the eye tracking system (for safety).
**Eye-Tracking System**
The eye tracking system used in the experiment was Tobii eye tracking device, where the hardware used was Tobii X2-60 Eye Tracker and the software used was Tobii Studio 3.3.
**Stimuli**
This pairwise group had eighteen diagrams in total. Each property had nine diagrams and the other one was redrawn keeping the same information but changing only the targeted curve. Each diagram had 6 curves which had 24-30 data items, as shown in Figure 2.
**Layout Guidelines**
We considered ten guidelines laid out by Blake et al. [2], [3] for drawing Euler diagrams. They are as follows[3]
Guide 1: Draw wellmatched Euler diagrams.
Guide 2: Draw wellformed Euler diagrams.
Guide 3: Draw smooth curved Euler diagrams.
Guide 4: Draw zone area equality Euler diagrams.
Guide 5: Draw diverging lines Euler diagrams.
Guide 6: Draw Euler diagrams without regard to orientation.
Guide 7: Draw circle shaped Euler diagrams.
Guide 8: Draw highly symmetrical curves Euler diagrams.
Guide 9: Draw shape discriminated Euler diagrams.
Guide 10: Draw Euler diagrams with curves that have no fill and different colours for each curve.
**Layout Characteristics**
To minimise unwanted variations between diagrams, we adhere to the drawing conventions and principles listed below, also considered by Blake et al. [2], [3].
1.All diagrams were drawn using 3 types of curve sizes: small, medium and large.
2.Diagrams were drawn with circles.
3.Diagrams were drawn with smooth curves and considered zone area equality.
4.Curve colours were picked from the colour palette shown in [3] To help ensure the colours in the palette were uniformly distinct from each other, we adopted the approach of Blake et al. [3].

---

[2] Confirmed by Ishihara test and a Snellen chart.

[3] A detailed explanation can be found in [2] and [3].

5. Curves were positioned such that the set properties like containment, disjointness and intersection were considered.
6. Curve labels were written using upper case letters in Times New Roman, 12 point size, font in bold as shown in Figure 2.
7. Each curve label was positioned closest to its corresponding curve either on top or bottom or right aligned as shown in Figure 2.
8. Data items were written using Times New Roman, 12 point size (started with capital letter).
9. Data items were evenly distributed within the zone. In some cases, data items were placed close to the curves while testing the properties.
10. Curves were arranged adhering to the layout guidelines considered above.

Further drawing conventions considered in the study are consistent with other research contributions [1], [2], [3], [4].

**Study Tasks**

We considered the set-theoretic concepts like set intersection, set inclusion and set disjointness. The tasks used in the study were based on students taking modules on a University degree course. The examples are shown in Figure 1. This concept is very familiar to all participants who took part in this study.

- How many students have taken SCIENCE but not TOURISM?
- Which module is being taken by 6 students?
- Who is taking both SCIENCE and TOURISM?

**Software**

We used the software called as the 'research vehicle' developed by Andrew Blake at the University of Brighton, United Kingdom [1-3]. This software can display the diagrams along with questions and options to the participants and records the time taken to answer the questions and the errors committed by the participants. The Tobii eye tracking software records the full session of the research vehicle for each participant.

**Independent Variables**

The three main independent variables used in the study are diagrams, tasks and participants.

**Dependent Variables**

The three main dependent variables used in the study are accuracy of answers, time taken to answer each question and eye movements.

**Procedure**

The experiment had three sessions:

**First session** - during this session, consents were taken from the participants and the purpose of the experiment was explained in detail. Participants were trained with different diagrams and task questions manually and using the software. Only after the participant was comfortable with the questions and diagrams, he or she could take up the next session.

**Second session** - in this session, participants were told to take up the study using the software; approximately lasting around 12 minutes where accuracy, time and eye movements were recorded. All the questions were supposed to be answered by the participants.

**Third session** - participants gave feedback about the questionnaire and the diagrams shown in the A4 paper (preference of diagrams).

**EXPERIMENTAL RESULTS**

In this study, the main interest is on eye movements of the participants. Given the small number of participants and the limited variety of the stimuli, we consider accuracy and response time as an additional evidence in support of eye movement findings.

**Accuracy**

From the task performance analysis (chi-square test), there is no significant difference ($p>0.05$) in accuracy between disconnected zones and duplicated curve labels, although the number of errors committed by participants are more in the latter.

**Response Time**

From the task performance analysis (ANOVA test), there is no significant difference ($p>0.05$) in response times between disconnected zones and duplicated curve labels, although average time taken by the participants is more in the latter.

**Eye Movements**

So, from the task performance analysis for accuracy and response times, it is clear that there is no significant difference between the properties. For this reason, eye movements of participants are important for the analysis.

From the video analysis, the pair-wise diagram (disconnected zones versus duplicated curve labels) was analysed. Participants' overall eye movements were slow (to some extent) for diagrams with duplicated curve labels. There were quite a lot of back-and-forth eye movements around the duplicated curves. This indicates that the participants were uncertain about counting the number of data items. Also, participants claimed that they were affected by the duplicated curves.

**Preference**

Eight out of twelve participants preferred disconnected zones over duplicated curve labels and the reason is because of the continuity of the curve (single curve). Since the curve is coloured, participants feel it is easy to follow the curve path rather scanning the full diagram to find the data items. Whereas, in duplicated curve labels, participants feel they spend more time in scanning the full diagram to find the duplicated curves and their data items.

**DISCUSSION**

In this controlled study experiment, parameters such as participants, diagrams and tasks can be a threat to our results validity but based on our empirical evidence through eye tracking, we established one of the principle for presenting an Euler diagram effectively in real world scenarios.

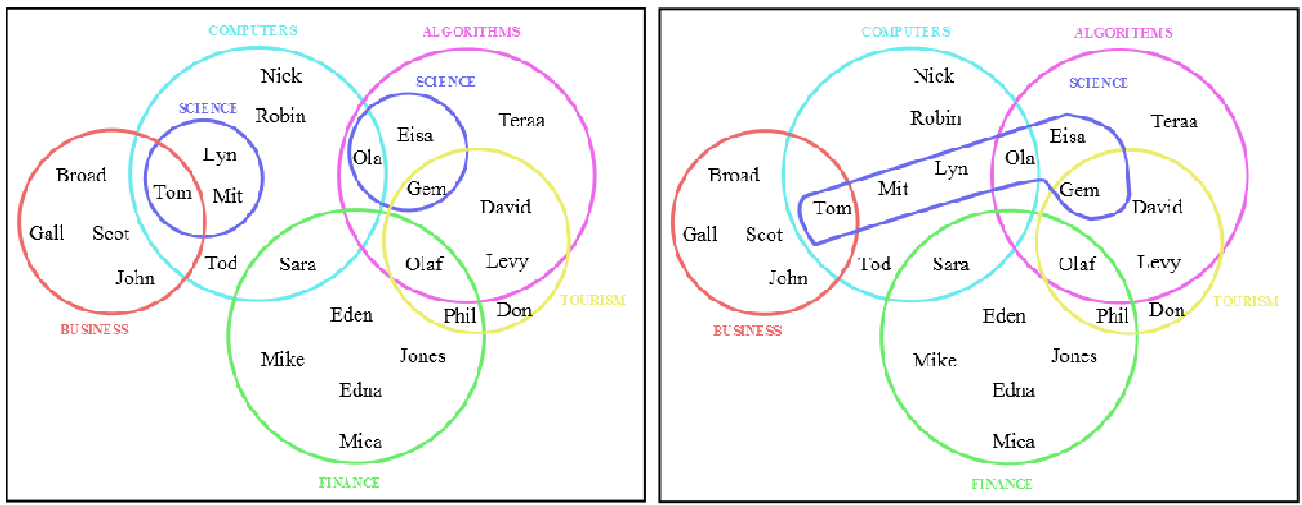

**Figure 2. Diagrams with (a) duplicated curve labels (SCIENCE) (b) non-duplicated curve label**

Many diagrams cannot be drawn without breaking one or more wellformedness property and from Rodgers et al. [7] we know there are different choices between the properties while generating Euler diagrams and from the study we conclude saying, duplicated curves should be avoided where possible because the eye tracking has revealed duplicated curve labels can be a barrier to understanding. The alternative is to use an irregular curve that may or may not disconnect a zone but there is continuity in the curve rather duplication. The results are summarised below:

1. From the accuracy results, though there is no significant difference between duplicated curve labels and disconnected zones.
2. From the response time results, there is again no significant difference between the two properties.
3. From the eye tracking results, most of the participants took more time for the duplicated curves. Participants' overall eye movements were slow for diagrams with duplicated curve labels. There were quite a lot of back-and-forth eye movements around the duplicated curves. This indicates that the participants were uncertain about counting the number of data items. Also, participants claimed that they were affected by the duplicated curves.
4. From the preferential results, 70% of the participants preferred disconnected zones over duplicated curve labels and the reason is because of the continuity of the curve (single curve).

From gestalt principles on similarity, similar curves with shape, colour and label should be grouped together for improving human understanding. If the duplicated curves are extreme far apart, there is a tendency of human missing out on one of the duplicated curve. Whereas, in the case of irregular non-duplicated curve label, again based on gestalt principles on closure: simple closed curves improves human understanding. From the Gestalt psychology on understanding diagrams, ``mind understands external stimuli as a complete object rather than the sum of their parts''. So, we conclude saying, avoiding duplicated curves will improve human understanding.